\documentclass{JHEP}
\usepackage{epsfig}

\newcommand{\be}{\begin{equation}}
\newcommand{\ee}{\end{equation}}
\newcommand{\bea}{\begin{eqnarray}}
\newcommand{\eea}{\end{eqnarray}}

\newcommand{\pq}{$(p,q)~$}

\preprint{TAUP-2608-2000\\
{\tt hep-th/0002118}}

\title{On the Spatial Structure of Monopoles}

\author{Barak Kol and
Michael Kroyter
\\
School of Physics and Astronomy\\ Tel Aviv University\\ Ramat Aviv 69978,
Israel\\
E-mail: \email{barak1k@post.tau.ac.il},\email{mikroyt@post.tau.ac.il}
}
\abstract{
We study the spatial structure of 1/4 BPS solitons in 4
dimensional ${\cal N}=4$ gauge theory. A weak binding
approximation is used where the soliton is made of several
``ingredient'' particles. Some spatial moduli are
described which are not accounted for in the (p,q) web
picture. These moduli are counted and their effect on the
solutions is demonstrated. The potential for off BPS
configurations is estimated by a simple expression and is
found to agree with previous expressions. We discuss the
fermionic zero modes of the solitons, and find agreement
with web predictions.}

\keywords{Supersymmetric Effective Theories, Solitons Monopoles and Instantons, D-branes}

\begin{document}

\section{Introduction}
We continue the study of the BPS spectrum of the maximally supersymmetric gauge
theory in 4 dimensions, namely ${\cal N}=4$. In the introduction we begin with a
review of our current knowledge, then we describe the open questions, and then
the contribution of this paper.


\subsection{Review of BPS particles in 4d ${\cal N}=4$ gauge theory}

The bosonic part of the Lagrangian of 4d ${\cal N}=4$
gauge theory can be written as
\be
  S = -{1\over 16\pi} \mbox{Im}
                  \int \tau ~Tr(F^2 - i F \wedge F)
        - {1\over 2g^2}\int ~Tr\Big( |D\phi^i|^2 +
     \sum_{i<j}[\phi^i,\phi^j]^2 \Big)
\ee
where $g$ is the coupling, $\tau=\theta/2\pi+i4\pi/g^2$ is
the complex coupling that incorporates also the theta
angle $\theta$, $F$ is the field strength and $\phi^i,
~i=1..6$ are scalar fields. All fields are in the adjoint
of some gauge group $G$. At a generic point in moduli
space the scalars acquire a VEV
$\vec{\phi}=diag(\vec{\phi_1},...,\vec{\phi_r})$ and the
gauge group is broken to $U(1)^r$, where $r={\rm
rank}(G)$. We consider states carrying various electric
and magnetic charges $(p,q)_i$ under these $U(1)$'s, and
we refer to all of them as `monopoles'.

Branes give a useful way to model this system. For gauge
group $SU(N_c)$ one takes $N_c$ parallel D3 branes, and
considers the scaling limit (``the field theory limit'')
$M_s \to \infty$ ($M_s$ is the string scale) keeping all
gauge theory energies fixed $E \sim \Delta x ~M_s^2$,
where $\Delta x$ is any (shrinking) length scale
perpendicular to the D3's.

In the case
$G=SU(2)$ the BPS spectrum is well known. It includes the $W$, the
monopole and in general all $(p,q)$ dyons. All the states are $1/2$
BPS and lie in a short (vector) multiplet. Actually, they are all
$SL(2, {\bf Z})$ duals of each other.

When we take a bigger gauge group $1/4$ BPS states become possible. We
consider mainly $G=SU(3),SU(4)$. The mass of such states (when they
exist) is given by the BPS formula
\be
M[(p,q)_i]=|Z|=\sqrt{\vec{Q_e}^2+ ~\vec{Q_m}^2 + ~2|\vec{Q_e} \times
\vec{Q_m|}}~
\footnote{The $Q$'s are six dimensional vectors, yet we find it convenient to
use the $|\vec
{Q_e} \times \vec{Q_m}|$ to represent $|Q_e| |Q_m|\sin{\alpha}$, as it is done
in \cite{BergmanKol}.}
\ee
where $\vec{Q_e}+i\vec{Q_m}=\sum_{j=1}^{r}{(p_j+\tau ~q_j)\vec{\phi_j}}$.

It was shown that \pq strings \cite{AharonyHanany,Zwiebach,AharonyHananyKol,Sen}
are important tools in analyzing $1/4$ BPS monopoles \cite{Bergman,BergmanKol}
(for related work see \cite{Kumar:1999qx,Gauntlett:1999vc}). Recall that a
string web is a planar collection of strings in the $(x,y)$ plane each carrying
a \pq label (\pq are relatively prime integers) and satisfying
\begin{enumerate}
\item{ Slope. The slope of a \pq string is given by $\Delta x +i
    ~\Delta y \parallel p+ \tau q$}
\item{Junction. \pq strings can meet at vertices as long as the \pq
    charge is conserved: $\sum{p_i}=\sum{q_i}=0$}
\end{enumerate}

Any \pq web that can be drawn with external legs $(p,q)_j$ all ending on D3
branes is identified with a monopole carrying the electric and magnetic charges
$(p,q)_j$ under the $U(1)$'s corresponding to each D3. It was shown that the
mass of the web is the same as the BPS mass
 \cite{Bergman}.
Moreover, the web picture leads to predict that the monopole will reach {\it
marginal stability} when a junction coincides with a D3 brane \cite{Bergman}.
This prediction was verified by the classical solutions of \cite{LeeYi} who
found that the size of the solution diverges as marginal stability is
approached.

Knowing the mass of the monopoles, one would like to know the possible
spins, namely the
multiplet structure. Being $1/4$ BPS it must contain the medium
representation (with $|j| \le 3/2$) as a factor. In \cite{BergmanKol}
the maximum spin $j$ in the multiplet was predicted by counting the number
of fermionic zero modes (FZM) on the web:
\be
|j| \le F + n_X/2
\label{max_spin}
\ee
where $F=F(p,q)$ is the number of internal faces in the
corresponding web and $n_X$ is the number of external
legs. So far little is known from field theory about the
multiplets when internal faces are present. In a case with
multiple external legs the known data \cite{LeeYi} is
consistent with the conjecture (\ref{max_spin}).

The growth of the degeneracy $d=d(p,q)$ (multiplet size) for large
charges was discussed in
\cite{KolThermal}. There it was ``phenomenologically'' found that the
ground state entropy $S=\log{d}$, behaves like
\be
S \sim \sqrt{F}.
\label{entropy}
\ee
where $F$, the number of internal faces, is quadratic in the charges.

Several other related studies appeared
\cite{BLLY,Kol:1998,Kol:1999,Bak:1999,Gauntlett:1999,LeeYi:1999,
Bak:1999b,Tong:1999,Lee:1999,Bak:1999c},
including studies of the potential energy of classical configurations of 1/4 BPS
states.

\subsection{Open questions}
So far we discussed predictions for monopoles from the web model. It is
natural to proceed in two directions: one is to test the web predictions in
field theory and the other is to study monopole properties which are not modeled by
webs.

In the first category
we would like to know the exact multiplet structure, or at least
to perform tests of the maximum spin prediction (\ref{max_spin}) and the
ground state entropy (\ref{entropy}).

Here we implicitly assume the existence of ``large'' or ``accidental'' BPS
representations. In black hole physics a BPS black hole has a huge exact
degeneracy, but in field theory this is unfamiliar. One would like to test
whether the webs in $SU(3)$ with many faces are indeed made of a sum of SUSY
representations, and moreover that the planar $SU(4)$ monopoles are BPS although
they are in a large SUSY representation \cite{BergmanKol}.

In the second category, we would like to study the spatial structure of the
monopoles. The webs are thought to live in a point in the D3, and so do not give
direct spatial information (Nevertheless, recall that Nahm's equations for
monopoles can be obtained from brane configurations \cite{Diaconescu}). In this
paper we try to study this problem directly in the field theory, in a certain
convenient limit.

We would like to mention an alternative geometric approach
to the problem which involves the study of special
Lagrangian submanifolds of $A_n \times {\bf T}^2$. $A_n$
is a non-compact K3, a blow-up of an $A_n$ singularity and
$n=2,3$ are of special interest. This formulation arises
since compactifying type IIB on this manifold gives us the
relevant field theory with gauge group $A_n=SU(n+1)$, and
D3 branes which wrap the special Lagrangian cycles give us
the 1/4 BPS states. We would like to look at an arbitrary
3-homology class, and determine its moduli space with flat
connections. (The 3-cycles of $A_n \times {\bf T}^2$ are
spanned by products of a 2-cycle in $A_n$ with a 1-cycle
in ${\bf T}^2$. Here we consider an arbitrary linear
combination of these.) Actually, we are interested in the
cohomologies of this moduli space, and each cohomology
determines a state in the sought after multiplet. At the
moment special Lagrangian submanifolds are an active field
of research in mathematics \cite{Hitchin,Joyce,Mclean}
(see also \cite{JoyceRelated}), and the results may be
available soon.

\subsection{In this paper}
In this paper we will study the spatial structure of the 1/4 BPS
states both for its own sake and in order to study the open questions
described above from a new angle.

To analyze the spatial structure we restrict ourselves to
limiting configurations - weakly bound monopoles - as
explained in section \ref{weakbind}. Moreover, we look at
the effective low energy theory on one of the $U(1)$
factors, or equivalently on one of the D3 branes. Thus we
will approximate the theory by  4d ${\cal N}=4$ super
Maxwell theory, and at times we will refer also to the
full super Born-Infeld (BI) on the brane for comparison.

In \cite{CallanMaldacena} solutions representing a single string (of F or D
type) emanating from a D3 were studied in the framework of the full Born-Infeld
action of a D brane \cite{APS}. More elaborate configurations involving string
webs were studied in \cite{Gutowski:1999,GKMTZ} and
the BI BPS solutions were found to solve the Maxwell equations as well.
For other related results see \cite{Hashimoto:1998a}.
In solutions of
Maxwell theory all fields are harmonic, that is, are
linear combination of $1/|r-r_i|$ potentials. These solutions represent a
collection of the ingredient particles with some of their relative distances
being constrained. The locations of the singularities $r_i$ are moduli of the
solutions (after accounting for the constraints), which  are not evident in the
web picture. In section \ref{moduliSec} we compare these moduli with the moduli
of the web. We find that for webs with 4 external legs or more there are
essential spatial moduli which are not seen in the web picture.
It is interesting to check the effect of these moduli on the web plane.
We present graphs of this variation showing how it affects the thickness of
junctions. On the other hand, we did not find in this approximation the moduli
of the web which correspond to changing the size of an internal face.

In section \ref{energetics} we study the potential for off BPS configurations or
the restoring force when one tries to change a constrained relative distance of
the solution. We use two methods, the first uses the super Maxwell picture and
the other uses the web picture. In both cases we find the same restoring force.
We compare our results to the expression for the full potential found in
\cite{BLLY} and find agreement. Whereas the latter expression is more general
our expression is a useful simple approximation.

In section \ref{fzm} we study the zero modes (especially
the fermionic ones) of some solutions. First we analyze
the solutions corresponding to 1/2 BPS states, an F string
and a D string ending on a D3, and we write down the
fermionic zero modes (FZM) explicitly. Then we discuss the
FZM of a 1/4 BPS solution, where we find that the FZM
counting coincides with the expectations of
\cite{BergmanKol} for the case under study.

\section{Weakly bound monopoles}
\label{weakbind}

One way to make the spatial structure tractable is to study weakly
bound monopoles. In this case the bound state has a clear structure -
it is made of well separated particles (which are more elementary as we
will explain), and some of the relative geometry is fixed by the
dynamics.

For concreteness let us consider our prime and simplest example - the weakly
bound simple junction. The string web, fig.(\ref{fig:simple}) is made of a D
string and an F string which intersect in a junction to produce a light $(1,1)$
string. The fundamental string ends on the D3 brane denoted by A, the D string
ends on B and the $(1,1)$ on C. In field theoretic terms, it is the state in an
$SU(3)\to U(1)^2$ gauge group, with charges $(1,0)_A,~(0,1)_B,~(-1,-1)_C$ under
the three $U(1)$'s corresponding to the three D3's (of course there are only two
independent $U(1)$'s because of the center of mass constraint. This redundancy
is reflected in the constraint on the charges by having both the electric and
magnetic ones sum to zero).

\FIGURE{\epsfig{file=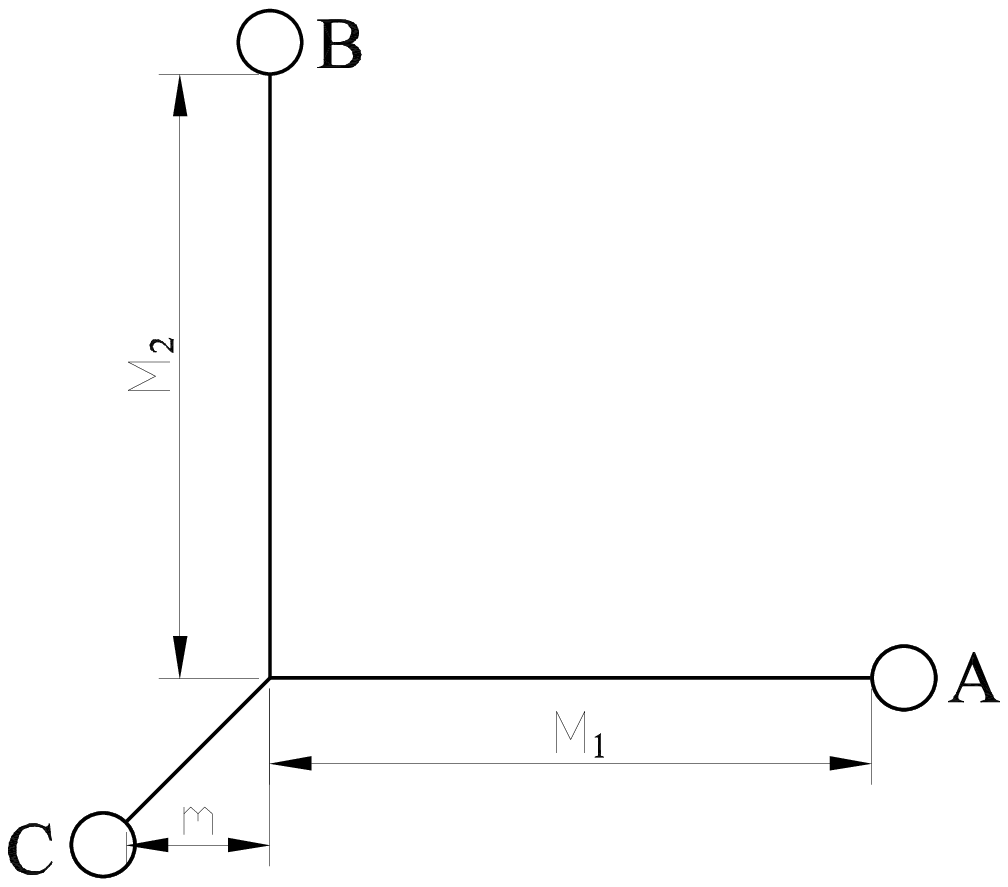,width=9.5cm}

\label{fig:simple}
\caption{The simple junction. The D3's are represented by circles.
One of the legs is short.}}

Assume for simplicity that $\tau=i$ (unit coupling), that the mass of the
fundamental string segment is $M_1$, that the mass of the D string segment is
$M_2$, and that the mass of the light $(1,1)$ string is $2m, ~m \ll M_1,M_2$.
This state has the same charges as a magnetic monopole in the $U(1)_{BC}$ (the
$U(1)$ which corresponds to relative BC motion), and an electric W particle in
$U(1)_{AC}$,, and hence could be thought to be a bound state of the two. The
binding energy is
\bea
E_b &= &~M_W+M_{mon}- ~M_{junc} ~= \nonumber  \\
    &= &\sqrt{(M_1+m)^2+m^2}+\sqrt{(M_2+m)^2+m^2} -(M_1+M_2+2m)
    ~=\nonumber  \\
  &=&m^2 ~/2\mu_M + O(m^3 ~/M^2)
\label{E_bind}
\eea
where $1/\mu_M=1/M_1+1/M_2$ is the reduced mass.

At low energies the effective action is that of a $U(1)^2, ~{\cal
 N }=4$ gauge theory. We choose to concentrate on the $U(1)$ which
lives on the short leg (we can do that since the different $U(1)$'s
are decoupled at low energies). We are interested in solutions in
which two of the scalar fields $X,Y$ are excited, in addition to the
gauge field.

As a general rule, short distances in the web plane are translated into large
distance in the D3 world volume (the UV-IR relation). To see this in the
case of the simple junction \cite{LeeYi,GKMTZ}, note that since the scalar fields
must satisfy the
Laplace equation (with sources) they should be of the form
\bea
X={1 \over |\vec r-\vec r_1|} \nonumber \\
Y={1 \over |\vec r-\vec r_2|}
\eea
where $\vec r_1$ is the world-volume location of the
electric charge, while $\vec r_2$ is the location of the
magnetic one. We see that $|\vec r_1-\vec r_2|$ is
constrained by $m$ to be
\be
m=Y(\vec r_1)=X(\vec r_2)={1 \over |\vec r_1-\vec r_2|},
\label{mInvr}
\ee
thus establishing the inverse proportionality.


\section{Soliton moduli from several approaches}
\label{moduliSec}

The solutions found in \cite{CallanMaldacena,GKMTZ} are interpreted as representing
strings and string junctions.
To get a better understanding of this correspondence, we would like to
compare the moduli of both. We will find that there are moduli which
are found on the weak binding approximation side
but not in the web picture and vice versa.

For the $SU(2)$ gauge group considered in \cite{CallanMaldacena} there are no
moduli
other than translations.
The same is true in the web picture.

\subsection{$SU(3)$}
For $SU(3)$ the simplest web is the string junction (see
figure (\ref{fig:simple},\ref{fig:su3})). In the weak
binding approximation that we use here, we will choose the
location of one of the D3's at the origin, while the other
two are at $(m,\infty)$ and $(\infty,m)$. These
asymptotics fix some of the moduli. Note that $m$ is not a
modulus of the solution, rather it is determined by the
field theory VEVs, since we consider the space of all webs
which terminate on a given configuration of D3's.

 To represent this
configuration in the language of the effective action we
have to consider solutions of the Laplace equation for
both $X$ and $Y$ which obey the asymptotics. The general
form of a solution with $n$ singular points is:
\be
\label{XYeq}
X=\sum_{i=1}^n{p_i \over |\vec r-\vec r_i|} ~~~~~~~~~~~~~~~~
Y=\sum_{i=1}^n{q_i \over |\vec r-\vec r_i|}
\ee
where the $\vec r_i$ are the spatial D3 coordinates. In addition to
the scalar fields there is also a vector
field, whose field strength is given by
\be
\label{EBeq}
\vec E=\vec \nabla X ~~~~~~~~~~~~~~~~ \vec B=\vec \nabla Y
\ee
We see that $p_i,~q_i$ are electric and magnetic charges, and as such must be
quantized in the quantum theory. Quantization reduces the $SL(2,{\bf R})$
symmetry of these solutions to $SL(2,{\bf Z})$. To get two semi-infinite strings
(In addition to the short one) we have to consider a two-centered solution. To
get the desirable charges we take $p_i=\{1,0\}$, $q_i=\{0,1\}$. When $X
\rightarrow \infty$ we get $Y \rightarrow m \equiv {1 \over |\vec r_1-\vec
r_2|}$, and for $Y \rightarrow \infty$ we get $X \rightarrow m$. By fixing the
boundary condition $m$, we fix the value of ${1 \over |\vec r_1-\vec r_2|}$
(\ref{mInvr}). This is the only parameter we have. Classically the other
parameters may be set to any value by affine transformations of the D3. In the
quantum theory these moduli would be quantized and we assume that the S wave
will be supersymmetric. We get that there are no essential moduli in this side
either.

\FIGURE{
\epsfig{figure=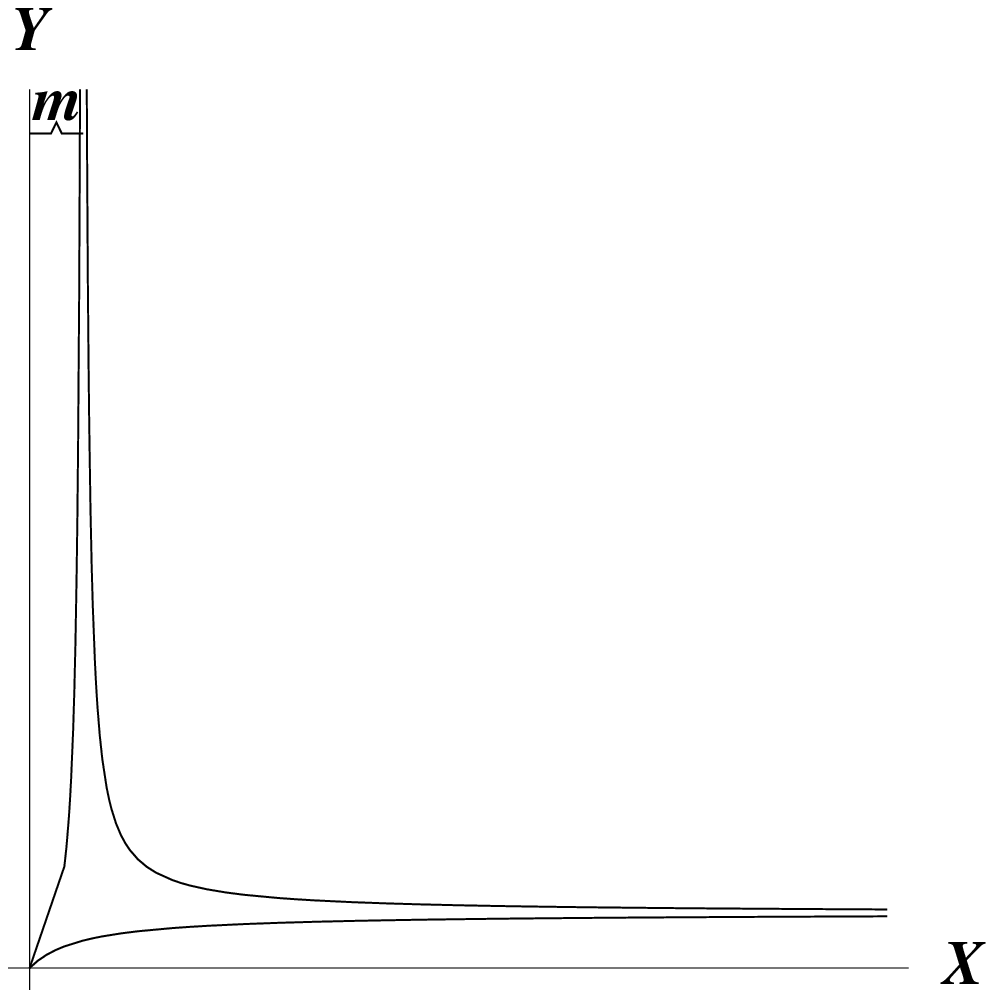,width=5.8cm}
\caption{The basic $SU(3)$ solution.}
\label{fig:su3}}

We show here the projection to the \mbox{$X-Y$} plane of
this configuration, fig.(\ref{fig:su3}). Note that the D3
brane surface though planar in this projection, is not
planar in the ten dimensional sense, since $X \rightarrow
\infty$ for $\vec r \rightarrow \vec r_1$, whereas $Y
\rightarrow
\infty$ for $\vec r \rightarrow \vec r_2$. Note also that
the length scale in the \mbox{$X-Y$} plane is inversely
proportional to the length scale in the D3 (\ref{mInvr}).

Whenever a grid diagram \cite{AharonyHananyKol}, which
corresponds to a given web has an inner point, there
exists a modulus that corresponds to ``blowing up a hidden
face'' \cite{AharonyHananyKol}. The simplest such
configuration for the $SU(3)$ case is represented in
fig.(\ref{fig:su3grid}) (the grid), and
fig.(\ref{fig:su3web}) (the web). Note that the length
parameter $a$ which appears in the web figure is not coded
in the grid. $a$ is a modulus of the configuration since
changing it does not change the mass of the web.

We would like to argue that such an internal face cannot
exist in the $U(1)$ field theoretical description. The
scalars map the ${\bf R}^3-\{points\}$ worldvolume into
${\bf R}^2$, the \mbox{$X-Y$} plane. By a hidden face we
actually mean not only an incontractible loop in the
target space, but an incontractible loop in the graph of
the map in $({\bf R}^3-\{points\}) \times {\bf R}^2$.
However, since $\pi_1({\bf R}^3-\{points\})=0$ that would
be impossible. It may still be possible, though, to
represent this modulus in field theory with higher gauge
groups.

\FIGURE{\epsfig{figure=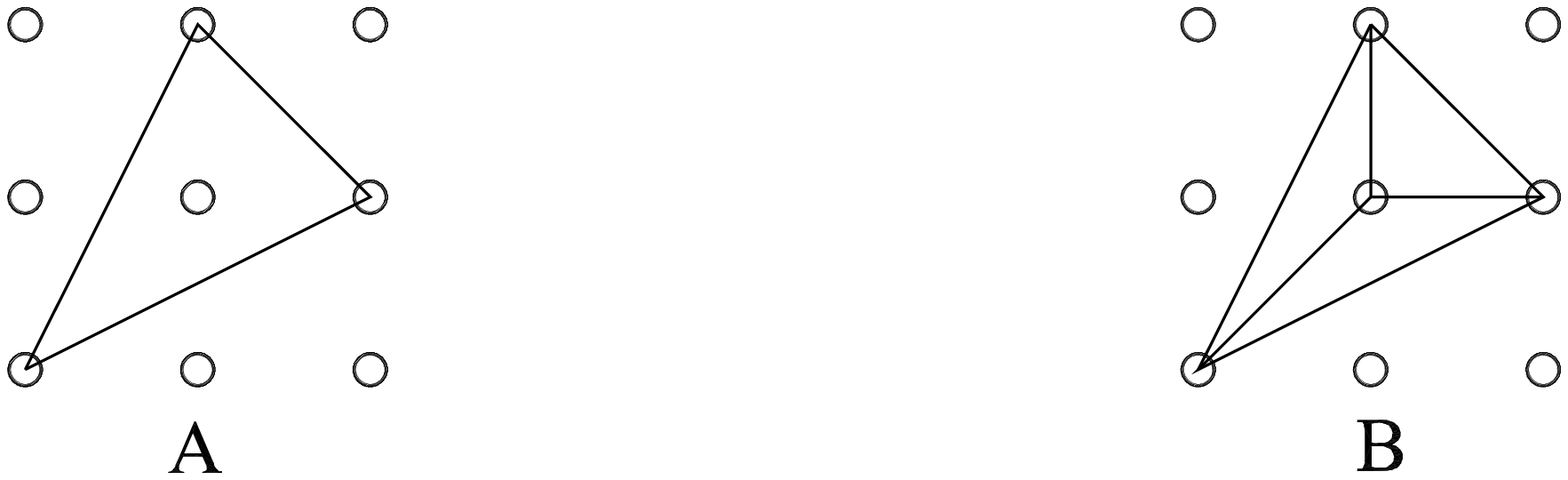,width=9cm}
\caption{The grid diagram of an $SU(3)$ solution.
(A) before, and (B) after the blowup.}
\label{fig:su3grid}}

\FIGURE{\epsfig{figure=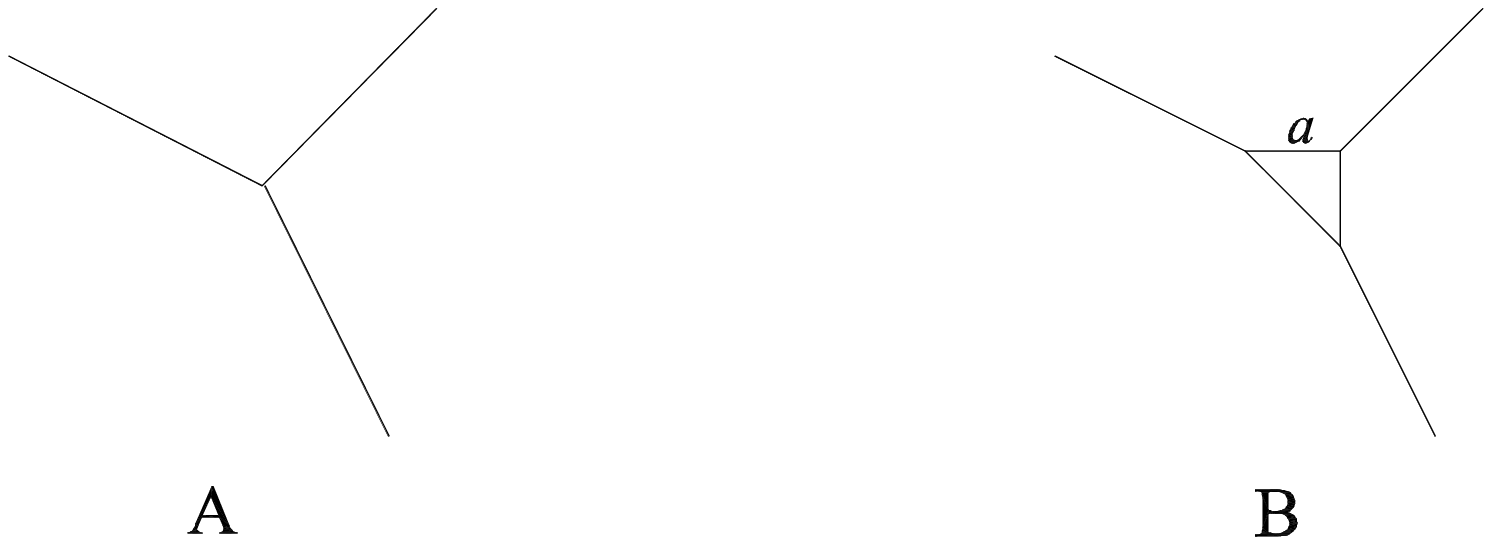,width=9cm}
\caption{The web diagram of an $SU(3)$ solution.
(A) before, and (B) after the blowup.}
\label{fig:su3web}}

\subsection{$SU(4)$}
We have seen a case where a modulus exists in the web
picture, but not in the ``weak binding'' approximation,
and a mild example to the opposite (mild in the sense that
the modulus did not affect the \mbox{$X-Y$} projection of
the configuration). To see an example where the weak
binding approximation yields an essential modulus
overlooked by the web we have to go to $SU(4)$. The
simplest $SU(4)$ configuration is given by $(1,0)$ and
$(0,1)$ strings joining to form a $(1,1)$ or a $(1,-1)$
string, which then splits again. It is easy to see that
there are no web moduli in this case. For the field theory
we have to choose the short leg first. We choose it to be
the left side $(1,0)$ leg. The solution is:

\be
X={1 \over |\vec r-\vec r_1|} ~~~~~~~~~~~~~~~~
Y={1 \over |\vec r-\vec r_2|}-{1 \over |\vec r-\vec r_3|}
\ee

We see that when $Y \rightarrow \infty$, $X \rightarrow a \equiv{1 \over |\vec
r_1-\vec r_2|}$, when $Y \rightarrow -\infty$, $X \rightarrow b\equiv{1 \over
|\vec r_1-\vec r_3|}$, and when  $X \rightarrow \infty$ $Y \rightarrow a-b$.
When $a-b$ changes sign, there is a transition from a $(1,1)$ internal leg to a
$(1,-1)$ internal leg in the corresponding web. In the grid diagram this
transition is represented by going from the grid of fig.(\ref{fig:su4grid}A) to
that of fig.(\ref{fig:su4grid}B).

\FIGURE{\epsfig{figure=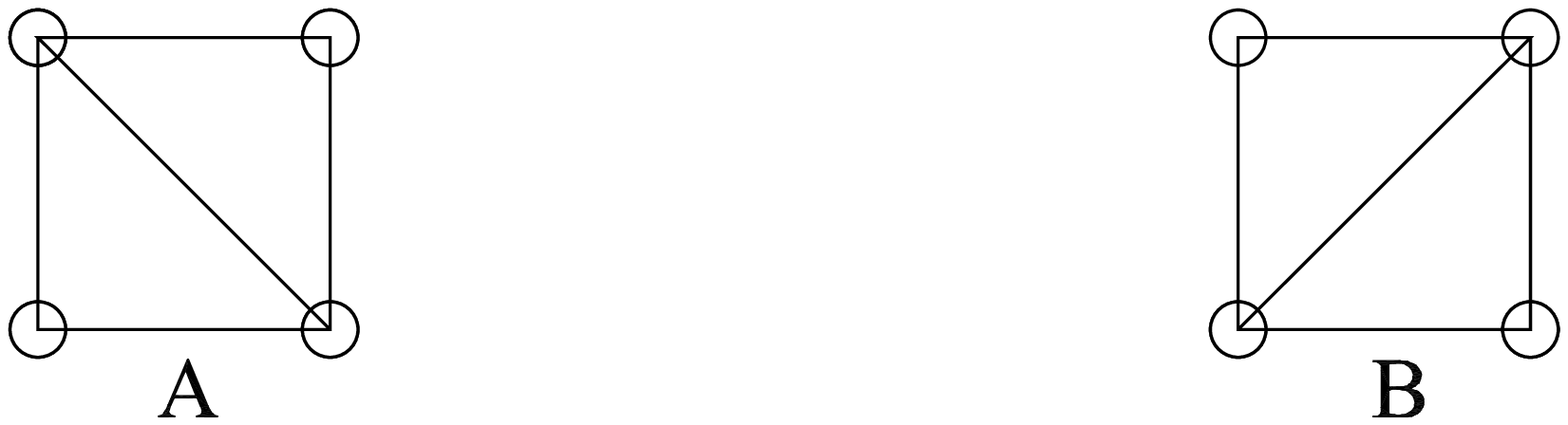,width=8.2cm}
\caption{The grid diagram of the $SU(4)$ solution.}
\label{fig:su4grid}}

In the D3 the three singular points define a plane, in
which $a$ and $b$ are two edges of a (possibly singular)
triangle. The location and orientation moduli do not
change the shape of the \mbox{$X-Y$} projection. Since $a$
and $b$ are fixed by the boundary conditions there is only
one modulus left, the angle between these two edges. It
is, nevertheless, one modulus more then in the web
picture. To see what is the meaning of this modulus we
simply show the projection of the configuration on the
\mbox{$X-Y$} plane for several values of the angle $\alpha$
between the two edges, fig.(\ref{fig:su4field}). It is
clear why this modulus is not visible in the web as our
web ``has no width''.

\FIGURE{\epsfig{figure=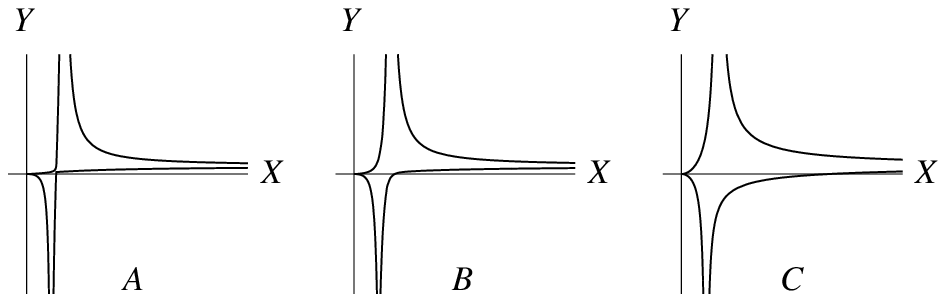,width=13cm}
\caption{The projection of the field theory solution for different values of
$\alpha$. In $A$, $\alpha=0$, in $B$, $\alpha={\pi \over 8}$ and in $C$,
$\alpha=\pi$.}
\label{fig:su4field}}

\subsection{Analogy with smooth membranes}
We will find some analogy between the moduli we find here
and moduli of complex curves describing a smooth M2
configuration. Such a description via a smooth M2 appears
after a compactification of the theory on a circle of
radius $L$. By doing that, though, we can no longer
discuss the four dimensional field theory which we had on
the D3. Therefore this analogy is only qualitative. We
continue the discussion with this in mind.

Following the analogous case of
\pq five branes in \cite{AharonyHananyKol} (based on \cite{Witten4M})
we take the coordinates to be
\be
s=\exp((X+ix_t)/L_t) ~~~~~~~~~~~~~~~~~~~~~ t=\exp((Y+iy_t)/L_t)
\ee
where $x_t$ and $y_t$ are the coordinates on the M-torus
of length $L_t$ ($\tau=i$). %
The equation
\be
F(s,t)=0
\ee
where $F$ is holomorphic, defines a surface $S$ in the
space $M={\bf R}^2\times{\bf T}^2$ parameterized by
$(X,Y,x_t,y_t)$

For the $SU(4)$ case above we read from the grid that $F$
should be the sum of four monomials: $1$, $s$, $t$ and
$st$. We can divide by the coefficient of $1$. The
coefficients of $s$ and $t$ determine the origin of the
axes, and can be scaled to $1$ as well. We are left with
the curve $F(s,t)=1+s+t+Ast$, where $A$ is a complex
coefficient. With this choice of coefficients the
asymptotic behavior of two out of the four legs is
determined to be $(X,Y)
\rightarrow (0,-\infty),(-\infty,0)$. Note that this is
possible, since in the M-theory picture we do not use the
one short leg approximation, but rather we consider all
legs to be semi-infinite with no D3's present. We are left
with one constrained parameter $a$, which describes the
asymptotic values of the other two legs $(X,Y) \rightarrow
(a,\infty),(\infty,a)$. It is easy to see, by considering
the asymptotic behavior of the other two legs that
\be
|A|=\exp(-a/L_t)
\ee
However, the argument of $A$ is not fixed. When, say, $X
\rightarrow \infty$, $x_t\rightarrow Const$. This constant
is represented by the argument of $A$. We see that like
the field theory, the M-theory representation has one
modulus. Projections of the curve on the \mbox{$X-Y$}
plane are represented in fig.(\ref{fig:M}) for several
values of this modulus.

\FIGURE{\epsfig{figure=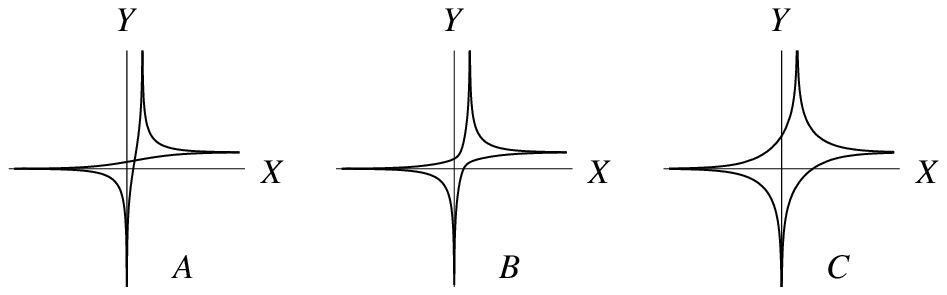,width=13cm}
\caption{The projection of the M-theory solution for different values of
$\alpha$. In $A$, $\alpha=0$, in $B$, $\alpha={\pi \over 8}$ and in $C$,
$\alpha=\pi$.}
\label{fig:M}}

There are some similarities between the field
configuration fig.(\ref{fig:su4field}) and the M-theory
solution fig.(\ref{fig:M}). In both cases the modulus is
an angle. Note also that in both cases the center of the
configuration expands fast, as figures
(\ref{fig:su4field}B,\ref{fig:M}B) correspond to $\alpha
= {\pi \over 8}$, rather than ${\pi \over 2}$.

It is possible to represent in the M-theory language all
of the web moduli. It would have been nice if all of the
weak binding moduli could be represented as well. The
simplest check one can perform is to compare the dimension
of the moduli space. We consider now webs which can be
represented by the field theory. These webs have no
``hidden faces'', and therefore the associated grid
diagram has no inner points. The total number of points in
this diagram, which is also the number of monomials in
$F$, is equal to the number of external legs. The absolute
values of all the coefficients is fixed by the boundary
condition. As for the arguments, we can again scale away
three of them. The number of moduli is
\be
\label{Mcount}
n_{M-moduli}=n_{Legs}-3
\ee
This number makes no sense for $n_{Legs}<3$, where the
number of moduli is just zero as there are no monomials to
scale.

On the weak binding side each leg, except the short one,
is produced by a singular point. Each such point can be
located anywhere on the D3, which gives $3(n_{Legs}-1)$
parameters. From this number we have to subtract the
number of constraints, which is $n_{Legs}-2$, since the
short leg is not counted, and the location of the last leg
is determined by the others. Then there is also the group
of affine transformation of the D3 which we should divide
by. We are left with
\be
n_{Field-moduli}=2 n_{Legs} -7
\label{field_moduli}
\ee

The last expression is valid only for $n_{Legs} \ge 4$,
while for $n_{Legs}=2,3$ we have $n_{Field-moduli}=0$.
This happens because not all of the $6$ affine parameters
are relevant. For $2$ legs there is one singular point, so
the rotations are immaterial, while for $3$ legs, there
are two points, so one rotation is irrelevant. For
$n_{Legs}=4$ both equations
(\ref{Mcount},\ref{field_moduli}) give us one modulus, as
shown above. For $n_{Legs}>4$ however, the equations
differ. Not only that, but the weak binding side has more
moduli then the M-theory side.

This difference can be formally accounted for by recalling
that instead of a D3, we have now a M2 which has one fewer
dimension. If we repeat the counting that we did for a
brane of dimension $p$, we get
\be
n_{Field-moduli}=p(n_{Legs}-1)-(n_{Legs}-2)-{p(p+1)
\over 2}
\ee
For $p=2$, we get exactly the same number as in (\ref{Mcount}).

\section{Energetics}
\label{energetics}

We have seen that in the weak binding limit a 1/4 BPS monopole is
composed of a number of constituents which are arranged spatially such
that they are at equilibrium. We shall now compute the restoring
potential - the potential for configurations which are close to
equilibrium.

The computation is carried out for the case of the simple
junction. As in fig.(\ref{fig:simple}) we have a short
$(1,1)$ leg of mass $2m$, and two long legs: one electric
oriented along the $x$ axis with mass $M_1$, and the other
magnetic oriented along the $y$ axis with mass $M_2$. It
is a bound state of an electric particle and a magnetic
particle at a distance $1/m$. As will be shown later, the
relative potential we get at large separations is
\be
V(r)={1 \over 2 \mu_M r} ~({1 \over r}-2m),
\label{potential}
\ee
where $r$ is the relative separation,
$1/\mu_M=1/M_1+1/M_2$ is the reduced mass and terms
subleading in $m/ \mu_M$ were neglected. It satisfies that
the equilibrium is at $r=1/m$ with binding energy $E_b=m^2
/(2
\mu_M)$, as expected (\ref{E_bind}). In addition we find
that the frequency of small oscillations is
\be
\omega ^2=m^4 ~/\mu_M^2.
\label{osc_freq}
\ee
Note that this frequency is of the same order as the
binding energy.

We use two different methods to derive (\ref{potential}).
One uses field theory and the other uses webs. Then we
successfully test it against the results of \cite{BLLY},
which gives an expression for the complete potential.
While our expression for the potential holds only close to
equilibrium it has the advantage of being simple in form
and derivation.

\subsection{Field theory computation}
The (static) force between the two particles is determined by their gauge
charges and scalar charges according to
\bea
F(r)=-{{p_{1i}p_{2i} + q_{1i}~q_{2i}} \over r^2}
 +{ {\Lambda_{1j}\Lambda_{2j}} \over r^2},
\label{force}
\eea
where \pq are electric and magnetic charges, $i$ runs over the
different gauge fields, $j$ runs over the different scalars, and
$r=|\vec r_1-\vec r_2|$. The
gauge charges are conserved and cannot depend on $r$, while the scalar
charges $\Lambda (r)$ change with $r$.

In order to find $\Lambda (r)$ recall that (by definition) we have
\bea
X = {\Lambda_{X1} \over |\vec r-\vec r_1|} +
 {\Lambda_{X2} \over |\vec r-\vec r_2|} \\
Y = {\Lambda_{Y1} \over |\vec r-\vec r_1|} +
 {\Lambda_{Y2} \over |\vec r-\vec r_2|}
\eea
At equilibrium $\Lambda_{X1}=\Lambda_{Y2}=1$ and
$\Lambda_{X2}=\Lambda_{Y1}=0$. In order to compute the lowest order
contribution to the force equation (\ref{force}) it is enough to
assume $\Lambda_{X1}=\Lambda_{Y2}=1$ for all $r$. We get two constraints by looking
at the scalar fields near the singularities at $r_1,r_2$ and requiring
that they pass through the D3 branes. Near $r_1$
\bea
M_1+m&=& X \simeq {1 \over |\vec r-\vec r_1|} +
 {\Lambda_{X2} \over r} \\
m&=&Y \simeq {\Lambda_{Y1} \over |\vec r-\vec r_1|} +
 {1 \over r}
\eea
from which we can solve for $\Lambda_{Y1}$. A similar argument near $r_2$
solves for $\Lambda_{X2}$
\bea
\Lambda_{X2}=(m- {1 \over r}) /M_2 \\
\Lambda_{Y1}=(m- {1 \over r}) /M_1
\eea

Now we substitute back in the force equation (\ref{force}) (there are
no gauge forces)
and then
integrate and find the potential to be exactly (\ref{potential}).

\subsection{Web computation}
It is interesting to note that a simple computation within the web
model can reproduce the result (\ref{potential}) as well. We know that
for separation $|\vec r_2-\vec r_1|=1/m$ the junction is at
equilibrium in coordinates $(X,Y)=(m,m)$.
We shall calculate the potential of a the radial mode, that is the
potential of the configurations for which the junction is at
$(X,Y)=(1/r,1/r)$.

Let us compute the mass of the web by summing the masses of all
strings and {\it neglecting} any interactions between them. After
subtracting the masses of the two particles we find a potential
\be
V(r)= \sqrt{(M_1+ \delta)^2+\delta^2} +
\sqrt{(M_2+ \delta)^2+\delta^2} +{2 \over r} -V_{\infty}
\ee
where $\delta=m-1/r, ~V_{\infty}=  \sqrt{(M_1+m)^2 +m^2} +
 \sqrt{(M_2+m)^2+m^2}$. This coincides with (\ref{potential}) to second
order in $\delta$.

We can calculate also the frequency of another mode of
oscillations, that is, $\delta X=-\delta Y$ rather then
$\delta X=\delta Y$. In this case, however,
$\omega^2_{~\delta X=-\delta Y}
\propto {\mu_M
\over m} \omega^2_{~\delta X=\delta Y}$ which is a much
higher frequency then the binding energy
(\ref{E_bind}),(\ref{osc_freq}). So at the quantum level
there would be no fluctuations in this direction.

\subsection{Another test}
Let us test our expression (\ref{potential}) against the result of
\cite{BLLY}. We will make a test which relies only on the functional
form of their result, so we will not need to compare the various
constants which they use, the sole exception to this is the
identification of our parameter $\mu_M$ with their parameter $\mu$,
since both are supposed to represent the reduced mass.

The functional form is
\bea
V_{BLLY}(r)&=& A^2 ~f(r)+ ~B^2 ~/f(r) ~~-V_{\infty} \nonumber \\
f(r)&=&1+ {1 \over {2\mu r}}, ~~V_{\infty}=A^2+B^2
\eea
By comparing the location of the minimum $r_{min}=1/m$ and the binding
energy eq.(\ref{E_bind}) we determine $A,B$
\bea
A^2&=&2\mu  \\
B^2&=&2\mu(1+{m \over 2\mu})^2
\eea
With this identification of the constants, and neglecting higher order terms of
the small parameters ${m \over \mu}$ and $m r-1$, the two potentials coincide.

\section{Zero modes}
\label{fzm}

In this section we find some of the zero modes of the
solutions. Both bosonic zero modes (BZMs) and fermionic
zero modes (FZMs) have a geometric interpretation (though
it is much more transparent in the BZM case). The FZMs are
relevant to the multiplet structure and spin of the
solutions. This link is carried out by quantizing the FZM
and BZM into a quantum mechanics on moduli space, and
looking for the degeneracy and spin of the ground states.
Such an analysis would allow us to test the existence of
``accidental long representations'' and other predictions
in \cite{BergmanKol,KolThermal}. Here we will take some
steps towards finding these zero modes. We shall
concentrate on counting the FZMs, but before that we shall
discuss some BZMs for completeness.

\subsection{A comment on BI action and BPS states}
Monopoles are usually described by a field theory. In
order to compare the field theoretical results to the
brane picture one needs to consider the BI action, as was
done in \cite{CallanMaldacena}. Note that in general, the
effective action for several D3's is a non-Abelian
Born-Infeld action \cite{Tseytlin,Brecher,Myers}.

By taking the scaling limit, as we do here, the nonlinearities
can be neglected, and the theory becomes SYM.
A different limit is to consider
all legs in the web except for one to be infinite, in which case the non-Abelian
part may be neglected, and one gets the S-BI action
(equations (85)-(88) of \cite{APS})
\be
\label{BIAction}
S = - \int\sqrt{- {\rm det}\, (
 \eta_{\mu\nu} + F_{\mu\nu} + \partial_\mu \phi_\alpha
 \partial_\nu \phi^\alpha - 2 \bar\lambda (\Gamma_\mu + \Gamma_\alpha
 \partial_\mu \phi^\alpha)\partial_\nu \lambda
 + \bar\lambda \Gamma^m \partial_\mu
 \lambda \bar\lambda \Gamma_m \partial_\nu \lambda)}
\ee
where $\alpha=4..9$, and $m=0..9$
\footnote{\label{convFN}Our conventions are
the same as those of \cite{APS}. We do not decompose the ten dimensional spinors
to four dimensional ones. $\{\Gamma^m,\Gamma^n\}=2\eta^{mn}$,
where $\eta=(-+...+)$.
Both type {\bf IIB} spinors have positive chirality ($ Q=\Gamma^{11}Q $).
The gauge is such that $r_\mu=\phi_\mu$ for $\mu=0..3$ leaving the usual
six scalars. In the following we shall denote $X=\phi_4$ and
$Y=\phi_5$. The other four scalars would not be excited.}.
In this work we took both
limits, thereby getting the S-Maxwell action which we used throughout this
paper.

Moreover, as far as BPS states are concerned, there is no
need to consider the BI action anyway. In
\cite{CallanMaldacena,GKMTZ} it was shown that some BPS
solutions satisfy both the Maxwell and the BI equations of
motion. This is a general property. In addition to $\vec
E,~\vec B$ one can consider in the BI theory the fields
$\vec D,~\vec H$, which are defined by
\be
\vec D={\partial L \over \partial \vec E} ~~~~~~~~~~~~~~~~~~~~
\vec H=-{\partial L \over \partial \vec B}
\ee
Discarding the fermions in eq.(\ref{BIAction}),
eq.(\ref{XYeq},\ref{EBeq}) together with
\be
\vec D=\vec E ~~~~~~~~~~~~~~~~~~~~
\vec B=\vec H
\ee
can be shown to be BPS solutions, and so the BPS states of
Maxwell and BI theory indeed coincide. Evaluation of the
Lagrangian (\ref{BIAction}) at the BPS states gives
\be
\label{lag}
L_{BPS}=-(1+B^2)
\ee
while the energy density is
\be
\label{ham}
H_{BPS}=1+E^2+B^2
\ee
Note that the energy density simplifies to a sum of three
terms - the tension of the D3, the electric and the
magnetic energies.

On the other hand, to study non-BPS states, or linear
waves on a given BPS background, one has to use the
nonlinearities. The quadratic Maxwell action looks the
same evaluated at any background, thus suggesting that all
configurations have the same zero modes, once the
locations of the singular points are prescribed.

By using the S-BI action we would see different properties of the D3 theory than the ones
grasped by the SYM action. Although the conclusions of this section would not necessarily
be relevant to the description of the FZMs of the SYM monopoles, they would
be relevant to the description of the D3.

We must remember that as we are using approximations to
the full theory, such as S-Maxwell or BI, a FZM of the
approximate theory will be a good approximation only away
from the soliton, and may diverge in its vicinity even if
the full FZM does not. In addition these theories may
contain spurious solutions as well.

\subsection{Bosonic zero modes}
Before we turn to the more complicated task of finding
general FZMs, we discuss briefly some BZMs of the F-string
solution, namely BZMs associated with transverse scalars
- scalars which are not excited in the background. In this case the
equations (\ref{XYeq},\ref{EBeq}) defining the background
reduce to
\be
\label{FbackG}
X={1 \over |\vec r-\vec r_0|} ~~~~~~~~~~~~~~~~
\vec E=\vec \nabla X
\ee

We exclusively examine the BZMs of the transverse scalars.
In \cite{CallanMaldacena} the linearized equation of the
radial mode of a transverse scalar field was found to be
\be
-(1+{1 \over r^4})\partial_t^2\phi +
r^{-2}\partial_r(r^2\partial_r\phi)=0
\ee
to get the radial zero mode equation one has to drop the
time dependence out of this equation. One gets the radial
part of the Laplace equation, and it can be checked that
the angular dependence of the BZM equation is restored by
using the full Laplacian
\be
\nabla^2 \phi=0
\ee

One would like to find a normalizable solution to this
equation, but there are none. Solutions of the Laplace
equation are characterized by their angular momentum. Each
value of angular momentum $l$ has two types of solutions.
One is proportional to $r^l$, and the other to $r^{-l-1}$.
For $l=0$ we have the constant solution which represents
the motion of the brane as a whole in the $\phi$ direction
(a VEV for the $\phi$ field), and the $\phi={1 \over r}$
solution with the geometric interpretation of a rotation
in the $X-\phi$ plane.

We would be interested in the localized modes
($r^{-l-1}$), which represent a zero mode of the soliton
rather than a zero mode of the D3 brane. However, even the
localized $l=0$ mode is too singular. It is not a ``length
normalizable'' mode - a mode which has a chance to become
a normalizable zero mode in the full theory. This term
will be defined later. The modes with higher $l$ are even
more singular.

\subsection{Fermionic zero modes}

For any soliton
FZMs can be found by operating on them
with broken supersymmetries
\be
\label{brokenQ}Q_b({\bf Soliton})={\bf FZM}
\ee
We can produce FZMs from BZMs
by acting on them with preserved supersymmetries
\be
\label{preservedQ}Q_p({\bf BZM})={\bf FZM}
\ee
and vice versa
\be
Q_p({\bf FZM})={\bf BZM}
\ee
There is, however, no guarantee that all the FZMs will be
found in any of these ways. In the case where more FZMs
are present there are ``accidental'' large BPS
representations, as in the case of the planar $SU(4)$ web
\cite{BergmanKol}.

The general FZMs can be found by solving the FZM equation, that is, by solving
the linearized equation of motion of the fermions with time derivatives set to
zero. In order to get this linearized equation we have to
expand the Lagrangian up to the second order with respect to the fermions
around the solution.
Neglecting the last term in the square root of eq.(\ref{BIAction}),
we notice that it depends on the fermions only through the expression
\be
c_{\mu\nu}=
\bar \lambda (\Gamma_\mu+\Gamma_\alpha\partial_\mu\phi^\alpha)
  \partial_\nu\lambda ~~~~
\ee
The linearized Lagrangian is therefore
\be
\label{linLag}
L_l=
\left.
{\partial L \over \partial c_{\mu\nu}}
   \right \vert_{background} c_{\mu\nu}=L_s+L_t
\ee
where $L_s$ is the part of the Lagrangian which contains the spatial dependence,
and thus is relevant for finding the FZMs,
and $L_t$ is the part with the time dependence.
A direct calculation shows that (recall our conventions
from footnote \ref{convFN})
\bea
\label{L_s}
L_s &=& \bar\lambda(\vec \Gamma+(\vec E+ \vec B \times \vec E)(\Gamma^4-\Gamma^0) +
  \vec B \Gamma^5+\vec B \times \vec \Gamma) \cdot \vec \nabla \lambda  \\
\label{L_t}
L_t &=& \bar\lambda(\vec \Gamma \cdot (\vec E+\vec E \times \vec B)+
E^2 \Gamma^4+\vec E \cdot \vec B \Gamma^5-(1+E^2+B^2)\Gamma^0
)\dot\lambda
\eea
We shall use $L_t$ when we would discuss the normalizability of the FZMs.

The FZM equation is just the spatial part of the equation of motion derived from
$L_l$, that is, it is the equation one would get by variation of $L_s$
(recall eq.(\ref{XYeq},\ref{EBeq}))
\be
\label{FZMeq}
(\vec \Gamma+(\vec E+ \vec B \times \vec E)(\Gamma^4-\Gamma^0) +
  \vec B \Gamma^5+\vec B \times \vec \Gamma) \cdot \vec \nabla \lambda=0
\ee

\subsubsection{FZM of the F-string}
We start with the F-string solution eq.(\ref{FbackG}), after which
the more complicated configurations will be dealt.
It was shown in
\cite{LeePeetThorlacius} that this solution is supersymmetric.
It breaks eight out of the sixteen supersymmetries which are present in the D3
world volume theory.
The preserved/broken
supersymmetries of this solution were found to coincide with those of a F-string,
as it should be.
The broken ones are given by
\be
\label{cm_broken}Q_b=-\Gamma^{04}Q_b
\ee
We shall first find the eight FZMs generated by broken supersymmetries, and then we
shall solve the FZM equation.
We recognize these FZMs by using (\ref{brokenQ}) on the F-string background
(\ref{FbackG}).
\be
\label{EfzmSol}
\lambda=\vec E \cdot \vec \Gamma \epsilon =
  {(r-r_0)_i \over |\vec r-\vec r_0|^3} \Gamma^i\epsilon
\ee
where $\epsilon$ is a constant spinor obeying
eq.(\ref{cm_broken}). The standard S-Maxwell SUSY
variation gives the same result in this case as the S-BI
one \footnote{The S-BI variation is given here and in what
follows by eq.(84) of \cite{APS} after $\zeta^{(3)}$ is
calculated in our background. Note that the number of
supersymmetries in these equations is twice what we have.
This is so because it contains also the supersymmetries
which are broken by the D3.\label{susyvar}}.

We want to check whether the solution we found is
normalizable. For that we first remind briefly
the way in which zero modes should be dealt (see
\cite{Rajaraman} for details). The zero modes should be
elevated to the status of collective coordinates by giving
them time dependence. In the action we should set
\be
\lambda \rightarrow \sum_{a=1}^n \lambda_a(\vec r)b_a(t)
\ee
where $\lambda_a(\vec r)$ is the $a^{th}$ FZM, $b_a(t)$ are the new
collective coordinates, and $n$ counts all the zero modes ($n=8$ here).
Solutions of the FZM equation (\ref{FZMeq}) nullify $L_s$, the spatial part of the
action eq.(\ref{L_s}). From eq.(\ref{L_t}) we see that what remains is a quantum
mechanics of the collective coordinates
\be
S=\int b^\dagger_a M_{ab}\dot b_b dt
\ee
where the mass matrix ${\bf M}$ is defined by
\be
M_{ab}=\int \tilde L_t[\bar \lambda_a,\lambda_b] d^3r
\ee
where $\tilde L_t$ stands for $L_t$ with $\dot \lambda \rightarrow \lambda$.
We call a set of solutions normalizable if all the entries of the mass matrix are
finite.

In the F-string case the mass matrix is given by
\be
M_{ab}=\int \bar\epsilon_a(\vec \Gamma \cdot \vec E)
(\Gamma^0-\vec \Gamma \cdot \vec E
)(\vec \Gamma \cdot \vec E)\epsilon_b
d^3r
\ee
where the $\vec B$ dependent terms were discarded, and the
$E^2(\Gamma^4-\Gamma^0)$ term drops since $\epsilon$ obeys
eq.(\ref{cm_broken}). This can be simplified to
\be
M_{ab}=\int (\epsilon_a^\dagger\epsilon_b -
 \bar\epsilon_a \Gamma^i \epsilon_b E_i)E^2 d^3r
\ee
The second term in this expression vanishes. To show that we use
eq.(\ref{cm_broken}) again.
\be
\label{goodEq}
\bar\epsilon_a \Gamma^i \epsilon_b =
 \bar\epsilon_a \Gamma^{04} \Gamma^i \epsilon_b =
 \bar\epsilon_a \Gamma^i \Gamma^{04} \epsilon_b =
 -\bar\epsilon_a \Gamma^i \epsilon_b
\ee
We are left now with
\be
M_{ab}=\epsilon_a^\dagger\epsilon_b \int E^2 d^3r=
  \epsilon_a^\dagger\epsilon_b \int {2 \pi r^2 dr \over r^4}
\ee
which diverges. However, this divergence can be understood when we change
coordinates from $r$ on the D3 to $X$ on the string by
\be
X={1 \over r}
\ee
The divergent integral is proportional to
\be
\int  dX
\ee
This is a constant (smooth) density, and the divergence
comes only from the infinite length of the string. The
same divergence in fact is present in the energy integral
of our background \cite{CallanMaldacena}. In the full
theory, where we expect BPS solutions that represent
finite strings, modes similar to this one should be
present, and these modes would be normalizable. We shall
call modes with this degree of divergence
``length-normalizable'' (LN for short). We shall discard
modes with higher degree of divergence.

We found the FZMs which originate from the broken SUSY. We now
turn to solve the FZM equation (\ref{FZMeq}). In the F-string background
it reduces to
\be
(\vec \Gamma \cdot \vec \nabla-
(\Gamma_4-\Gamma_0){1 \over r^2} \partial_r  )\lambda=0
\label{FZMF}
\ee
where we have set $r_0=0$ for simplicity. Note that had we
used the Maxwell theory to obtain an FZM equation the last
term would be absent.

To find the solutions we define
\be
P={1 \over 2}(1+\Gamma_0 \Gamma_4) ~~~~~~~~~~ P'={1 \over 2}(1-\Gamma_0 \Gamma_4)
\ee
and decompose $\lambda$
\be
\lambda=\lambda_1+\lambda_2 ~~~~~~~~~~~~
 \lambda_1=P\lambda
 ~~~~~~~~~~~~
 \lambda_2=P'\lambda
\ee
The equation (\ref{FZMF}) becomes
\be
\vec \Gamma \cdot \vec \nabla \lambda_1=0 ~~~~~~~~~~~~~
\vec \Gamma \cdot \vec \nabla \lambda_2 =- {2 \over r^2} \partial_r
 \Gamma_0 \lambda_1
\ee
In the case $\lambda_1=0$ we get the equation
\be
\vec \Gamma \cdot \vec \nabla \lambda_2=0
\ee
Squaring the differential operator shows that $\lambda_2$ should be a
solution of the Laplace equation. We are interested in localized
solutions with singularity at the origin (at $\vec r_0$). We can write
\be
\lambda_2=\sum Y^l_m r^{(-l-1)} \lambda^l_m
\ee
with $\lambda^l_m$ constant spinors, and check which
conditions should these spinors obey.
It is easy to check that $l=0$ has no solution.
For $l=1$ the solutions are exactly what we have found above
(\ref{EfzmSol}).

For the case $\lambda_1\neq 0$ we will get for $\lambda_1$
the same equation that we have got for $\lambda_2$, namely
the three (spatial) dimensional free Dirac equation. The
only solution that we will consider is $l=1$, for which
$\lambda \propto {1 \over r^2}$. But now we have to take
this solution and substitute it in the equation of
$\lambda_2$. A solution of this equation, if exist at all,
would have to behave as ${1 \over r^4}$, which is too
singular. Thus we conclude that the SUSY generated FZMs
are the only LN FZMs in this case.

\subsubsection{FZM of the D-string}

For the D-string solution equations (\ref{XYeq},\ref{EBeq}) reduce to
\be
\label{DbackG}
Y={1 \over |\vec r-\vec r_1|} ~~~~~~~~~~~~~~~~
\vec B=\vec \nabla Y
\ee

Now the S-Maxwell and S-BI SUSY variations give different
expressions. The S-Maxwell gives a result very similar to
the electric one
\be
\lambda=\vec B\cdot \vec \Gamma\epsilon =
  {(r-r_1)_i \over |\vec r-\vec r_1|^3} \Gamma^i\epsilon
\label{FZMD}
\ee
where $\epsilon$ is a constant spinor obeying the equation
of a broken SUSY in the magnetic case
\be
\epsilon= - \Gamma^{1235} \epsilon
\ee
The S-BI SUSY variation on the other hand gives
\be
\lambda={\vec B \cdot \vec \Gamma+B^2 \Gamma^{123} \over 1+B^2}
  \epsilon
\ee
At a neighborhood of $\vec r_1$ this solution has a finite limit.
Far from the core the BI mode reduces to the Maxwell one.

This finite behavior, and the fact that the most singular term in $L_t$
goes like $B^2$ implies that these solutions are LN.
A direct calculation shows that they are
non-normalizable. In fact, the mass matrix here is identical
in form to that of the F-string
\be
M_{ab}=\epsilon_a^\dagger\epsilon_b \int B^2 d^3r
\ee

Note also, that while the expression we got by using the
S-Maxwell SUSY variation (\ref{FZMD}) does not solve the
S-BI FZM equation (\ref{FZMeq}), the same expression {\it
is} a solution after reversing the ``chirality'',
$\epsilon= \Gamma^{1235} \epsilon$. However, these
solutions are non-LN. In the planar case we shall meet
similar solutions which would be LN.

\subsubsection{FZM of a planar configuration}

In the planar case both electric and magnetic charges are present.
SUSY is preserved by supercharges obeying
\be
{1 \over 2}(1+\Gamma^{04})Q_{pp}=Q_{pp} ~~~~~~~~~~~
   {1 \over 2}(1+\Gamma^{1235})Q_{pp}=Q_{pp}
\ee
There are three sectors of broken SUSY which we label by
$Q_{pb},~Q_{bp},~Q_{bb}$, according to the sector which breaks SUSY
($[\Gamma^{04},\Gamma^{1235}]=0$).
For example $Q_{bp}$ breaks the electric and preserve the magnetic
SUSY, that is
\be
{1 \over 2}(1-\Gamma^{04})Q_{bp}=Q_{bp} ~~~~~~~~~~~
   {1 \over 2}(1+\Gamma^{1235})Q_{bp}=Q_{bp}
\ee

Using eq.(\ref{EBeq}) and the formulas of \cite{APS}
(recall footnote (\ref{susyvar})) we get for these three
sectors
\bea
\label{lamPB}
\lambda_{pb}&=& {\Gamma^0 (\vec \Gamma \cdot \vec B + B^2 \Gamma^{123})
  \over 1+B^2} \epsilon_{pb} \\
\label{lamBP}
\lambda_{bp}&=& {\vec E \cdot
  ((\vec \Gamma \times \vec B - \vec \Gamma)\Gamma^{123} + \vec B )
\over 1+B^2} \epsilon_{bp} \\
\label{lamBB}
\lambda_{bb}&=& {\Gamma^0 (\vec \Gamma \cdot \vec B + B^2 \Gamma^{123})
  + \vec E \cdot
  ((\vec \Gamma \times \vec B + \vec \Gamma)\Gamma^{123} - \vec B )
  \over 1+B^2} \epsilon_{bb}
\eea
The $\epsilon$'s are constant spinors. Each sector has
four independent spinors, which amounts to twelve FZMs.
Note that $\lambda_{\alpha \beta}$ does not have the same
eigenvalues with respect to $\Gamma^{04}$, $\Gamma^{1235}$
as $\epsilon_{\alpha \beta}$. For example
$\Gamma^{04}\epsilon_{pb}=\epsilon_{pb}$, but
$\Gamma^{04}\lambda_{pb}=-\lambda_{pb}$, and
 it is not an eigenvector of $\Gamma^{1235}$ at all.

One can verify that these modes are indeed solutions of
the FZM equation (\ref{FZMeq}). To check that they are LN
we note that at the vicinity of magnetic singularities, or
more generally any \pq charge for $q \neq 0$, these
solutions approach a constant. These singularities will
not cause a problem. We have to check the behavior of the
modes in the vicinity of the purely electric
singularities. The four modes (\ref{lamPB}) are
independent of $E$ and so only the other eight modes
(\ref{lamBP},\ref{lamBB}) are potentially problematic.
Computing their mass matrix and using eq.(\ref{cm_broken})
as in eq.(\ref{goodEq}), one can see that these modes are
LN.

These are not all the LN solutions of the FZM equation
(\ref{FZMeq}).
 Consider
\be
{(r-r_a)_i \over |\vec r-\vec r_a|^3} \Gamma^i \epsilon_{bp}
\ee
where $\vec r_a$ is any singular point with no magnetic
charge, we see that it is a LN solution. The number of
these solutions is $4n_E$ where $n_E$ is the number of
electric singular points. However, since these modes exist
only for configurations with this kind of singular points,
they are probably artifacts of our approximation, since
they are not symmetric with respect to electric-magnetic
duality.

Next we note that (\ref{lamBP}) is linear with respect to
$\vec E$. Replacing $E_i$ in (\ref{lamBP}) by ${(r-r_a)_i
\over |\vec r-\vec r_a|^3}$, where now $\vec r_a$ is any
singular point, with or without magnetic charge, we get
another LN solution to the FZM equation (\ref{FZMeq}). A
web with $n_X$ external legs will have together with the
modes of (\ref{lamPB},\ref{lamBB}) $4n_X$ FZMs
\footnote{Recall that each such `mode' actually represents four modes,
and note that by replacing $\vec E$ by $\vec B$, (\ref{lamBP})
is reduced to (\ref{lamPB}).}.
This exactly coincides with
\be
n_{FZM}=8F+4n_X
\ee
of \cite{BergmanKol} for the number of web FZMs, for the
case $F=0$ of no internal faces. As we mentioned in
section \ref{moduliSec} we can not describe configurations
with internal faces in the effective action language. In
particular we see that in the $SU(4)$ case there are
indeed $16$ FZMs, in agreement with  \cite{BergmanKol}.

We did not show that the solutions we found are all the LN solutions.
However, we considered an ansatz, similar in form to the
solutions we found $\lambda={numerator \over 1+B^2}\epsilon$,
where the numerator is a sum of terms at most linear with respect
to ${(r-r_a)_i \over |\vec r-\vec r_a|^3}$ for any singular point times
factors at most quadratic with respect to the magnetic field.
For the $SU(3)$ and $SU(4)$ cases the solutions we have found
are the only ones of this form.


\acknowledgments{We thank Alon Marcus, Emanuel
Diaconescu, David Morrison, and Ehud Schreiber for discussions.\\ We thank the
organizers of the Jerusalem winter school and the Tel Aviv TMR string workshop
for a stimulating environment.\\ Research supported in part by the US-Israeli
Binational Science Foundation, the German--Israeli Foundation for Scientific
Research (GIF), by the European Commission TMR programme ERBFMRX--CT96--0045,
and the Israel Science Foundation.}

\end{document}